\documentclass[MSNbibl,nameyear,seceqn,dvips]{arxstspdf}
\usepackage{flushend}
\usepackage{stfloats}
\usepackage{graphicx}
\volume{29}
\issue{1}
\pubyear{2014}
\firstpage{50}
\lastpage{57}
\doi{10.1214/13-STS432} %kopijuoti is PTS
\makeatletter

\newcommand{\ga}{\gtrsim}
\newcommand{\unit}[1]{{\mbox{ #1}}}
\def\citealt{\cite}
\renewcommand{\citep}[1]{(\citeauthor{#1}, \citeyear{#1})}
\makeatother

\begin{document}
\begin{frontmatter}

\title{Finding the Most Distant Quasars Using Bayesian Selection
Methods}%\thanksref{T1}
% kai straipsnis turi susijusiu diskusiju ir rejoinder'iu
%rejoinder at \relateddoi{r}{10.1214/00-STSXXXX}.}
%
\runtitle{Finding the Most Distant Quasars}

\begin{aug}
\author[a]{\fnms{Daniel} \snm{Mortlock}\corref{}\ead[label=e1]{mortlock@ic.ac.uk}}
\runauthor{D. Mortlock}

\affiliation{Imperial College London}

\address[a]{Daniel Mortlock is Lecturer, Department of Mathematics,
Imperial College London,
London SW7 2AZ, United Kingdom, and
Astrophysics Group,
Blackett Laboratory,
Prince Consort Road,
Imperial College London,
London SW7 2AZ, United Kingdom \printead{e1}.}

\end{aug}

% ABSTRACT
%
\begin{abstract}
Quasars, the brightly glowing disks of material that can form around the
super-massive black holes at the centres of large galaxies,
are amongst the most luminous astronomical objects known
and so can be seen at great distances.
The most distant known quasars are seen as they were when the
Universe was less than a billion years old
(i.e., $\sim\!7\%$ of its current age).
Such distant quasars are, however, very rare,
and so are difficult to distinguish from the billions of
other comparably-bright sources in the night sky.
In searching for the most distant quasars in a recent
astronomical sky survey
(the UKIRT Infrared Deep Sky Survey, UKIDSS),
there were $\sim\!10^3$ apparently plausible candidates
for each expected quasar, far too many to reobserve
with other telescopes.
The solution to this problem was to apply Bayesian model
comparison, making models of the quasar population and the
dominant contaminating population (Galactic stars)
to utilise the information content in the survey measurements.
The result was an extremely efficient selection procedure
that was used to quickly identify the most promising UKIDSS candidates,
one of which was subsequently confirmed as the
most distant quasar known to date.
\end{abstract}

% KEYWORDS
% Pirmas kwd is didziosios raides
%
\begin{keyword}
\kwd{Astrostatistics}
\kwd{Bayesian methods}
\kwd{model comparison}
\kwd{astronomy}\vspace*{12pt}
\end{keyword}
\end{frontmatter}

%s1 #&#
\section{Introduction}
\label{section:intro}

Quasars
(e.g., \citealt{Rees:1984})
are amongst the most extraordinary astronomical objects known,
the result of material falling into the huge black holes
that lie at the centre of most galaxies (including the Milky Way).
Galaxies' central black holes typically
have masses of $M_\mathrm{BH}\ga 10^6 M_\odot$
(where $M_\odot= 1.99 \times10^{30} \unit{kg}$ is the mass of the Sun);
but these
black holes are mostly quiescent
(as is currently the case in the Milky Way).
Galaxies' central black holes only grow appreciably
when there is a gravitational disturbance,
possibly from other nearby galaxies,
that forces gas and stars off their otherwise stable orbits.
In the process of falling into the black hole,
this disturbed material becomes compressed,
forming a dense accretion disk
which is heated to extreme temperatures of
$\ga10^4 \unit{K}$.
Even though such accretion disks are typically only a billionth the size
of their host galaxies
(having radii more comparable to that of the Solar System),
their thermal glow can be brighter than all the stars in a galaxy,
and the brightest quasars
have luminosities of $L \simeq10^{15} L_\odot$
(where $L_\odot= 3.84 \times10^{26} \unit{W}$ is the
total power output of the Sun).
At cosmological distances it is often the
case that the host galaxy is too faint to be seen,
leaving only the bright point of light that is the glowing accretion
disk---it is these sources that are called quasars.\footnote{``Quasar''
is a contraction of ``quasi-stellar object,'' the name given to these
astronomical sources before their true nature was understood.
The fact that quasars,
like stars, appear as unresolved point-sources is one reason they
can be difficult to identify.}

As a result of their high luminosities,
quasars have
been amongst the most distant astronomical objects known
ever since their discovery \citep{Schmidt:1963}.
Due to the finite speed of light,
they are also
seen as they were when the Universe was much younger,
and \mbox{distant} quasars (DQs) can be used to reveal the
conditions in the early Universe.
At present, the most distant known quasars
(e.g., \citealt{Fanetal:2006c}, \citealt{Willottetal:2010}, \citealt{Mortlocketal:2011d})
are so
far away that they are seen as they were when the
Universe was less than~$1$ billion years old
(i.e., less than 7\% of its current age of
$13.8 \unit{billion years}$).
The remainder of this article only
concerns DQs in the
first billion years of the Universe;
although millions of more nearby quasars have been catalogued,
it is only the most distant (or otherwise unusual) that are important
as single objects.

%f1 #&#
\begin{figure}

\includegraphics{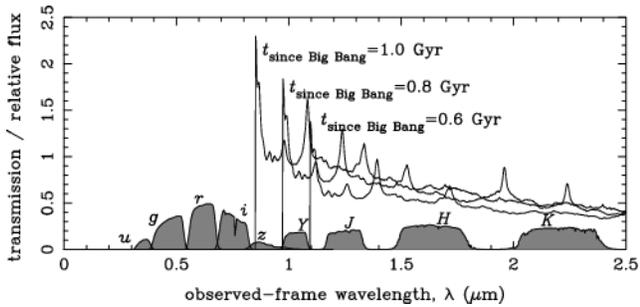}

\caption{Simulated spectra of DQs compared to the transmission curves of the SDSS optical filters
($u$, $g$, $r$, $i$, $z$)
and the UKIDSS near-infrared filters
($Y$, $J$, $H$, $K$).
The three quasars are at different distances and hence
different look-back times, as labelled.
The normalisation is arbitrary, although the decrease
in flux level with distance is realistic.
The least distant of these quasars might be seen in
SDSS images (specifically the $z$ band),
but the most distant would not.
All three simulated quasars would be visible in
appropriately deep UKIDSS exposures.}
\label{figure:spectrum}% \caption{}\label{}
\end{figure}

No current telescopes can resolve DQs as anything other than point-sources,
but
they are sufficiently bright that they can be observed
spectroscopically, with the light separated into different wavelengths.
As illustrated in Figure~\ref{figure:spectrum}, such observations can reveal
both the constituents of the quasars themselves and the
properties of the material along the lines-of-sight to them
(e.g., \citealt{Fanetal:2006b}).
The most important example of this is the distinct break
at an observed-frame wavelength of $\sim\!1 \unit{$\mu$m}$
that can be seen in all three
simulated quasar spectra shown in Figure~\ref{figure:spectrum}.
This break comes about as the
light at shorter wavelengths
is absorbed by neutral hydrogen atoms that were present in the
early Universe. But by ${\sim}1 \unit{billion years}$ after the Big Bang
the hydrogen in the Universe had been almost completely ionised,
as the first generations of stars---and quasars---emitted
sufficient ultraviolet radiation to separate electrons from
protons. The rest-frame wavelength of the break is at
$0.1216 \unit{$\mu$m}$, but the wavelength of all light
is increased by the cosmological expansion;
the Universe is a factor of $\sim\!7$ larger now than it was when
it was $1 \unit{billion years}$ old,
and so the breaks in the spectra of
these quasars are seen at an
observed-frame wavelength of
${\sim}7 \times0.1216 \unit{$\mu$m} \simeq0.85 \unit{$\mu$m}$.
The spectra of older quasars, including the break,
are redshifted by even more, hence the systematic shift with age
seen in Figure~\ref{figure:spectrum}.
More importantly, it is exactly these observations of DQs that currently provide the most direct evidence that
the hydrogen in the early Universe was indeed neutral
(e.g., \citealt{Fanetal:2006b}).

There is hence a great incentive to find quasars
at ever greater distances.
Unfortunately,
such DQs are incredibly rare,
with a number density on the sky of only one per $\sim\!200\  {\mathrm{deg}^2}$
(which is $\sim\!1000$ times the solid angle subtended by the Moon).
This would not pose a problem if there were no other astronomical objects
in the night sky,
but quasars are
generally seen as unresolved point-sources,
and there are
many billions of other point-sources (mainly stars in the Milky Way galaxy)
of comparable flux\footnote{The
basic quantity of brightness measured by a telescope is
flux density, averaged over the range of wavelengths
or frequencies to which the
detector is sensitive.
Several different definitions of flux density are used,
but here it is defined as energy per unit area, per unit time,
per unit frequency. A standard astronomical unit for flux density,
used here, is the millijansky, denoted $\mu$Jy,
which is equal to $10^{-32} \unit{W} \unit{m}^{-2} \unit{Hz}^{-1}$.
Further, as all astronomical measurements involve
averaging over a finite range of wavelengths or frequencies,
the term ``flux'' is used here as a convenient shorthand for
flux density.}
in the night sky.
Hence, it would be impossible to distinguish a DQ (or any type of rare object) in a single astronomical image;
but
it is possible to make observations of the same region of
sky using different filters,
which can be combined to give an estimate of a source's
colour.\footnote{The negative logarithm of
the ratio of (measured) fluxes in different bands is referred to as colour
in astronomy.
One utility of this quantity is that a source's
colours are largely independent of its distance, so they provide
information about its intrinsic properties even if the
distance is not known.
Distances to astronomical objects are often unknown,
and in some sense the main aim of the Bayesian selection procedure
described in Section~\ref{section:bayes}
is to determine whether a source is
thousands of light years from Earth
or
billions of light years from Earth.}

A key fact behind all the quasar selection methods
considered here---both heuristic and Bayesian---is that DQs can be
differentiated from stars on the basis of their measured colours.
This is particularly true for DQs because
of the sharp break in their spectra described above---most astronomical sources exhibit spectra which
vary much more smoothly with wavelength.
Unfortunately,
observational noise and the sheer variety
of astronomical sources combine to
render colour-based discrimination imprecise,
hence the need for some thought to be put into the selection methods.
This is best illustrated by example, as in Section~\ref{section:data},
which describes the properties of a recent astronomical survey
that was, in part, designed to find quasars more
distant than any previously known.
However, it quickly became apparent that the simple
methods that had previously been used to select candidate
DQs would not work because the level of
contamination was too high,
and
so a Bayesian selection method was developed
(Section~\ref{section:selection}).
The results of this approach are then described in Section~\ref{section:results}.

%%%%%%%%%%%%%%%%%%%%%%%%%%%%%%%%%%%%%%%%%%%%%%%%%%%%%%%%%%%%%%%%%%%%%%%%%%%%%%%%

%s2 #&#
\section{Data}
\label{section:data}

The DQ searches described here are based on
images of the sky taken during large astronomical surveys.
This is currently the dominant mode of observational astronomy.
Rather than the traditional practice of using telescopes to obtain
bespoke measurements of objects of interest, a survey involves taking
images of a large, usually contiguous, region of sky and then using
automated image-processing techniques to identify
and characterise
the sources that have been detected.
A single exposure is sufficient to measure the positions and morphological
characteristics of the detected sources, but it is usual
(particularly in the optical and near-infrared bands considered here)
to obtain measurements in each of several filters so that
some information about the sources' colours is also obtained.
The DQ search discussed here was based on
optical (Section~\ref{section:sdss}) and near-infrared (Section~\ref{section:ukidss}) surveys of the same
area of sky,
the data from which were combined to give the full wavelength
coverage shown in Figure~\ref{figure:spectrum}.

%s2.1 #&#
\subsection{SDSS}
\label{section:sdss}

The Sloan Digital Sky Survey (SDSS; \citealt{Yorketal:2000})
is one of the most ambitious astronomical projects ever undertaken:
over a period of several years
a quarter of the sky was imaged in each of
the five optical filters (denoted $u$, $g$, $r$, $i$ and
$z$)
shown in Figure~\ref{figure:spectrum}.
These filters span the wavelength range from $\sim\!0.3 \unit{$\mu$m}$
to $\sim\!1.0 \unit{$\mu$m}$
and so a DQ would appear invisible in all but the longest
wavelength filter (i.e., the $z$ band).

One of the most important results of SDSS was the discovery of
quasars seen as they were when the Universe was
$\sim\!1 \unit{billion years}$ old.
These objects were identified
by searching for sources which are detected in the
$z$-band images but undetected in the $u$, $g$, $r$ and $i$ filters (\citeauthor{Fanetal:2001}, \citeyear{Fanetal:2001}, \citeyear{Fanetal:2003}).
However, more ancient quasars could not be discovered by an optical
survey such as SDSS because, as shown in Figure~\ref{figure:spectrum},
these sources are only visible at the longer near-infrared wavelengths.
If the SDSS were to be expanded to cover more of the sky, it
would detect more such DQs,
and a similar
but more sensitive optical survey would find fainter quasars of
the same age; but no optical survey can detect significantly
more distant objects---progress can only be made by observing
at longer wavelengths.

%s2.2 #&#
\subsection{UKIDSS}
\label{section:ukidss}

The UKIRT Infrared Deep Sky Survey (UKIDSS; \citealt{Lawrenceetal:2007})
is the largest astronomical survey ever undertaken at near-infrared wavelengths
(i.e., just longer than $1 \unit{$\mu$m}$).
UKIDSS consists of five sub-projects, including the
Large Area Survey (LAS),
which was designed specifically to be a near-infrared counterpart to SDSS.

Essentially all the area covered by the LAS observations
had already been observed by SDSS
and so the two surveys' catalogues
could be combined.
As the surveys also have comparable sensitivities
(in the sense that the majority of sources bright enough to be detected
in one survey are also detectable in the other),
the cross-matched catalogue provides measured
fluxes in the nine filters shown in Figure~\ref{figure:spectrum}.
These measurements are denoted
$\{\hat{F}_b\} =
\{
\hat{F}_u,
\hat{F}_g,
\hat{F}_r,
\hat{F}_i,
\hat{F}_z,
\hat{F}_Y,
\hat{F}_J,
\hat{F}_H,
\hat{F}_K
\}$,
where
$b$ is the index of the band and
the carat denotes that these are estimators
constructed from the raw data.
As such, while the true flux (in any band, $b$)
satisfies $F_b\geq0$,
it is possible that
$\hat{F}_b< 0$
due to the stochastic nature of the measurement process
(specifically, the subtraction of an uncertain background level from the
original images).
Equally important, all these measurements have
known uncertainties,
$\{\sigma_b\} =
\{
\sigma_u,
\sigma_g,
\sigma_r,
\sigma_i,
\sigma_z,
\sigma_Y,
\sigma_J,
\sigma_H,
\sigma_K
\}$, although these are not only
different in each band,
but also different in each image that makes up the survey.
Taken together, the availability of
$\{\hat{F}_b\}$
and
$\{\sigma_b\}$ means that it is possible to evaluate the
likelihood accurately for each detected source (see Section~\ref{section:bayes}).

From the point of view of a search for DQs,
the critical point is that UKIDSS makes measurements
at sufficiently long wavelengths that it could detect
quasars at greater distances---and hence look-back times---than optical surveys like SDSS.
This was one of the main science goals of the UKIDSS LAS.
While the
UKIDSS data is demonstrably of sufficient quality to detect any such
DQs that might have been in the area observed \citep{Dyeetal:2006},
actually distinguishing them from the millions of other sources
in the UKIDSS catalogue proved to be a more challenging problem
than was initially expected.

%%%%%%%%%%%%%%%%%%%%%%%%%%%%%%%%%%%%%%%%%%%%%%%%%%%%%%%%%%%%%%%%%%%%%%%%%%%%%%%%

%s3 #&#
\section{Candidate Selection}
\label{section:selection}

The cross-matched UKIDSS--SDSS sample described above
includes $\sim\!2 \times10^7$
catalogued sources over $\sim\!5.5 \%$ of the sky;
given previous measurements of the target DQ population,
the expectation was that $\sim\!10$ of these would be
quasars in the first billion years of the Universe,
with maybe one or two beyond the distance limit that the SDSS could reach.
The first major step in the search process is to
select a set of candidates that is reasonably complete
(i.e., contains most or all of the DQs in the initial sample)
but small enough that follow-up measurements can be made to
obtain decisive classifications.
These reobservations can be either photometry
(requiring less telescope time) or
spectroscopy
(to provide a decisive classification),
but in either case it is only feasible to follow-up
a few hundred sources at most.
Hence, the initial catalogue must be reduced by a factor of
$\sim\!10^{-5}$ on the basis of the available
UKIDSS--SDSS survey data alone.
The numbers alone make this a challenging data-mining problem;
the complications of real astronomical data only make a difficult
situation worse.

%f2 #&#
\begin{figure*}

\includegraphics{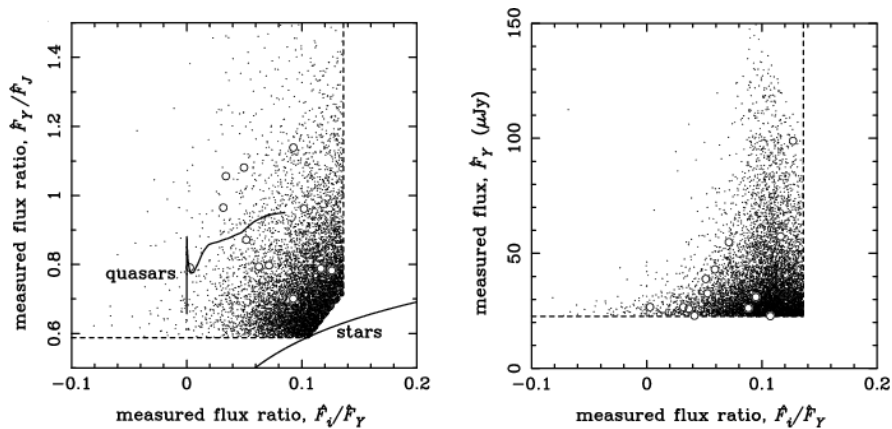}

\caption{Observed photometric properties of the
$\sim\!10^4$ sources in the UKIDSS--SDSS survey which have measured colours
(in the $i$,  $Y$ and $J$ bands)
consistent with being DQs.
The 13 large symbols are the confirmed DQs.
The dashed lines indicate the data cuts which were applied
before any more refined selection criteria
were applied.
The two solid curves in the left-hand panel illustrate the
ranges of intrinsic colours expected of DQs and stars.}
\label{figure:candidates}\vspace*{3pt}
\end{figure*}

The initial approach to this selection
problem was to focus on the astronomical
considerations, with little regard to any underlying statistical principles.
Any
catalogued source that
was completely inconsistent with being a DQ was discarded.
This included the following:
sources with colours in the
$i$, $z$, $Y$and $J$ bands that had values close to
zero (and hence no break);
objects whose images were visibly extended
(as evidenced by various image summary statistics calculated during
the automated data analysis process);
and sources that were sufficiently close on the
sky to other, brighter sources that their measured fluxes were
likely to be compromised.
This process, while complicated in detail,
is unremarkable conceptually,
and previous searches for DQs (e.g., \citealt{Warrenetal:1994}, \citealt{Fanetal:2001}, \citealt{Willottetal:2007})
included a similar series of steps to obtain a clean sample
of astronomical sources with reliable measured fluxes.
The result of applying these heuristic,
but automated, tests to the UKIDSS--SDSS data
was the sample of $\sim\!10^4$ catalogued sources
shown, along with the colour and flux cuts, in Figure~\ref{figure:candidates}.

All these sources have measured
fluxes
that are broadly like those expected for DQs,
but most are also obviously consistent with being Milky Way stars;
as is all too clear from Figure~\ref{figure:candidates},
the two populations overlap in the UKIDSS--SDSS data space.
In the previous DQ searches mentioned above
the quasars' colours were far more distinct,
with the result that
simple colour cuts
generated sufficiently small candidate samples
for follow-up observations.
In the case of the UKIDSS--SDSS search,
however,
there was a need
for some means of further
prioritising the candidates using only the
initial survey data.

%s3.1 #&#
\subsection{Candidate Ranking Schemes}

It clear from the data in Figure~\ref{figure:candidates}
that not all the candidates which passed the basic colour
cuts are equally promising.
Most of the candidates are clustered next to the selection
boundaries to the bottom-right of both plots,
whereas the fraction of DQs increases towards the top-left,
where there are also fewer sources.
This distribution is a convolution of the Galactic stars' instrinsic
colours with the measurement noise.
This is why the
spread in
$\hat{F}_i / \hat{F}_Y$
is broader for fainter sources
(i.e., towards the bottom of the right panel of Figure~\ref{figure:candidates}):
fainter objects' colours are measured with less precision.
Hence, an apparently dramatic---and promisingly low---value of
$\hat{F}_i / \hat{F}_Y$ is far less significant if, for example,
$\hat{F}_Y \simeq30 \unit{$\mu$Jy}$ than if, for example,
$\hat{F}_Y \simeq100 \unit{$\mu$Jy}$.
It is not sufficient merely for an object to have the
measured colours of a DQ;
its photometry must also be \textit{inconsistent} with it being a
Galactic star.
Indeed, the latter is in some ways a more stringent requirement
for a good candidate, given that there are so many more stars than
quasars.
Combining these considerations gives a clear,
if qualitative, scheme for prioritising the candidates:
identify those which are extreme outliers from the stellar
distribution but are also consistent with being DQs.
The remaining challenge is
to encode these ideas in a quantitative algorithm.

If the properties of the DQ and star populations were both
known perfectly, then Bayesian methods would be optimal.
But given that any highly ranked candidates would
be reobserved to confirm their nature,
there is no reason to prefer a principled statistical approach---it is only the relative ranking of the candidates which is important.
Any method which systematically separated
DQs from Galactic stars would suffice.
Many options are possible, but all must include
some measure of how likely it is that a source has been drawn
from the dominant stellar population.
An approach such as kernel density estimation would
have some of the desired properties, but the
distribution shown in Figure~\ref{figure:candidates} only appears
well sampled because it is seen in projection:
there is valuable information in the other measured fluxes
(particularly the $z$ band).
It is also problematic that most methods
of inferring the density of the stars' measured properties
would effectively involve a convolution, broadening the distribution;
the stars outnumber the DQs by such a large
factor that even a small broadening could have a drastic effect.
One way around this would be to instead attempt deconvolution,
to instead infer the true distribution of star colours,
to which noise could be added.
However, the only sensible way to utilise this estimate of
the stellar population would be as the population prior in the Bayesian
calculation suggested above.
The inevitable conclusion is that
all potential ranking methods would either involve the loss
of information (and, specifically, classifying power)
or be equivalent to the Bayesian approach.

%%%%%%%%%%%%%%%%%%%%%%%%%%%%%%%%%%%%%%%%%%%%%%%%%%%%%%%%%%%%%%%%%%%%%%%%%%%%%%%%

%s3.2 #&#
\subsection{Bayesian Candidate Probabilities}
\label{section:bayes}

The Bayesian quasar selection algorithm
is based on the calculation of the
probability, $P_{\mathrm{q}}$, that a candidate is a DQ (as opposed to a Galactic star).
This probability retains all the relevant information in the data
(i.e., including the uncertainties),
provided only that the models of the
quasar and star populations are accurate.
The various heuristic ranking schemes hinted at above are
hence incorporated naturally,
and the only arbitrary aspect of the resultant candidate
list is the minimum probability chosen for follow-up.

The starting point for the calculation of the candidate
probabilities is, of course, Bayes's theorem,
in this case applied to two hypotheses:\footnote{As discussed in detail by
\cite{Mortlocketal:2012a}, there are many other potential explanations
for an apparently promising candidate, including asteroids, supernovae
and a range of nonastronomical phenomena that relate to
problems with the detector or the data processing.
These phenomena are difficult to model
and easy to spot by visual inspection
and sub-dominant compared to the stellar contamination, so were
dealt with heuristically, rather than in a principled Bayesian way.}
that the candidate is a target DQ (${\mathrm{q}}$);
that the target is a Galactic star (${\mathrm{s}}$).
Given data on a source
in the form of measured fluxes $\{ \hat{F}_b\}$
in ${N_b}$ bands,
the probability that the source is a target quasar is
%
%e3.1 #&#
%
\begin{eqnarray}
\label{equation:pq} P_{\mathrm{q}} &=& \operatorname{Pr}\bigl({\mathrm{q}}| \{
\hat{F}_b\}, {\mathrm{det}}\bigr)
\nonumber
\\
&= &\bigl(\operatorname{Pr}({\mathrm{q}}| {\mathrm{det}}) {\mathrm
{Pr}}\bigl(\{ \hat{F}_b\}| {\mathrm{q}}, {\mathrm{det}}\bigr)\bigr)
\nonumber
\\[-8pt]
\\[-8pt]
\nonumber
&&{}\big/
\bigl(\operatorname{Pr}({\mathrm{q}}| {\mathrm{det}}) \operatorname{Pr}\bigl( \{ \hat
{F}_b\}| {\mathrm{q}}, {\mathrm{det}}\bigr)
\\
&&\hspace*{10pt}{}+ \operatorname{Pr}({\mathrm{s}}| {\mathrm{det}}) \operatorname{Pr}\bigl(\{ \hat
{F}_b\}| {\mathrm{s}}, {\mathrm{det}}\bigr)\bigr) ,\nonumber
\end{eqnarray}
where $\operatorname{Pr}({\mathrm{q}}| {\mathrm{det}})$ is
the prior probability
that the source is a quasar,
$\operatorname{Pr}({\mathrm{s}}) = 1 - \operatorname{Pr}({\mathrm{q}}|
{\mathrm{det}})$
is the prior probability that it is a star,
and $\operatorname{Pr}(\{ \hat{F}_b\}| {\mathrm{q}})$ and ${\mathrm
{Pr}}(\{ \hat{F}_b\}| {\mathrm{s}})$ are
the model-averaged likelihoods under each hypothesis.
Here ``{$\mathrm{det}$}'' indicates that all these probabilities are
conditional on the fact that the source has been detected at all.

The requirement that only detected sources are considered is a
critical and somewhat subtle ingredient in this formalism.
It ensures that
the prior distribution of each population's parameters
can be normalised unambiguously,
while avoiding the meaningless notion of an
unconditional prior probability of the nature of a source.
Asked out of context, the question
``What is the probability that a source is a DQ?''\vadjust{\goodbreak}
is ill-posed and
has no sensible answer.
This immediately implies that it is impossible to determine
the prior probability of a source being of a certain type without
at least some constraining information,
such as a range of fluxes or other observed properties.
Thus, the similar question ``What is the probability that a source
which has been detected by UKIDSS is a DQ?''
does have a well-defined answer,
the numerical value of which could be estimated
from the observed numbers of quasars and stars that
have fluxes which are greater than
the UKIDSS detection limit.

This would then be a reasonable empirical value for the quasar prior,
although even here the answer depends on
various other factors such as position in the sky.\footnote{If a source
is close on the sky to the Galactic plane
(visible even to the naked eye as the Milky Way),
then the number of contaminating stars above any flux limit
is greatly increased, which should be reflected in an
increase in $\operatorname{Pr}({\mathrm{s}}| {\mathrm{det}})$
and a resultant decrease in $\operatorname{Pr}({\mathrm{q}}| {\mathrm{det}})$.
(The quasar population does not vary significantly with
position on the sky because the target DQs are
at cosmological distances at which the distribution of all
objects is expected to be homogeneous and isotropic.)}
The implication
is that,
at least in principle,
the population priors would have to be calculated for each
source under consideration,
a potentially significant complication.
Fortunately,
the variation in, for example, the numbers of stars
with position on the sky within
UKIDSS is generally small compared to the
overall difference between the
numbers of quasars and stars.
This is illustrated by the fact that
of the $\sim\!10^7$ astronomical sources
catalogued by the UKIDSS LAS,
only $\sim\!10$ are expected to be DQs.
Hence, $\operatorname{Pr}({\mathrm{q}}| {\mathrm{det}}) \simeq10^{-6}$,
which might seem
an unusually extreme value for a prior, but is merely a reflection of
just how rare DQs are.

In the full implementation described in \citet{Mortlocketal:2012a},
empirical models for the star and quasar populations are
developed which give both the priors and a component of the likelihood.
Both populations exhibit a variety of observable properties,
most obviously because both DQs and
Galactic stars span a wide range of flux levels.
Denoting the $N_p$ parameters\footnote{For quasars the population
parameters are redshift and luminosity; for Galactic stars the
population parameters are colour (which serves as
an observable proxy for surface temperature) and luminosity.}
describing objects in population $p$
(which is either ${\mathrm{q}}$ or ${\mathrm{s}}$)
as
$\{ \theta_{p,i} \}
= \{ \theta_{p,1}, \theta_{p,2},
\ldots, \theta_{p, N_p} \}$,\
the model-averaged likelihood for population $p$ is
%
%e3.2 #&#
%e3.3 #&#
%
\begin{eqnarray}
\label{equation:evidence}&&  \quad \operatorname{Pr}\bigl(\{ \hat{F}_b\}| p, {
\mathrm{det}}\bigr)\nonumber
\\
&&\quad  \quad= \int\cdots\int \operatorname{Pr}\bigl( \{ \theta_{p,i} \} |
p\bigr)\operatorname{Pr}\bigl({\mathrm{det}}| \{ \theta_{p, i} \},
p\bigr)\\
&&\quad  \qquad {}\times\operatorname{Pr}\bigl(\{ \hat{F}_b\}| \{ \theta_{p, i} \}, p
\bigr)\,{\mathrm{d}}\theta_{p,1}{\mathrm{d}}\theta_{p,2}
\cdots {\mathrm{d}}\theta_{p,N_p},\nonumber
\end{eqnarray}
where $\operatorname{Pr}( \{ \theta_{p, i} \} | p)$
is the prior which describes the distribution of intrinsic
properties
of objects of type~$p$,
$\operatorname{Pr}({\mathrm{det}}| \{ \theta_{p, i} \}, p)$ is the probability
that an object with the specific properties would be detected,
and
$\operatorname{Pr}(\{ \hat{F}_b\}| \{ \theta_{p, i} \}, p)$
is the likelihood.

The detection probability is
calculated by approximating the source identification algorithm
of the UKIDSS LAS as being equivalent to a hard cut in the measured
flux that corresponds to a signal-to-noise ratio of 5
in the $Y$ and $J$ bands.
Fortunately, it is not critical to model the source detection
process accurately, as the other
heuristic astronomical cuts
(described at the start of Section~\ref{section:selection})
inevitably include effective flux limits as well.
The main role of this term is to ensure
that the populations' parameter priors have a clear normalisation.

The likelihood
must include information on how the parameters relate to observables,
as well as describing the
stochastic aspects of the measurement process.
In the case of optical or near-infrared survey photometry,
the dominant source of uncertainty is the additive
background noise in the image
(although the Poisson-distributed noise from the finite number of photons
received is important for the brightest sources).
The noise is approximately normal in flux units and,
neglecting inter-band correlations,
the likelihood
is taken to be
%
%e3.4 #&#
%
\begin{eqnarray}
\label{equation:likelihood} &&\operatorname{Pr}\bigl(\{ \hat{F}_b\}| \{
\theta_{p, i} \}, p\bigr) \nonumber\\
&&\quad= \prod_{b= 1}^{N_b}
\frac{1}{(2 \pi)^{1/2} \sigma_b}\\
&&\quad\qquad{}\times \exp \biggl( - \frac{1}{2} \biggl( \frac{\hat{F}_b- F_{p,b}
(\{\theta_{p,i}\})} {
\sigma_b}
\biggr)^2 \biggr),\nonumber
\end{eqnarray}
where
$F_{p,b}(\{\theta_{p,i}\})$ is the predicted flux
in band $b$
for a population $p$ source with properties
$\{ \theta_{p,i} \}$,
and $\sigma_b$ is the photometric uncertainty in band $b$.

This formalism for evaluating $P_{\mathrm{q}}$
is summarised in equations (\ref{equation:pq}),
(\ref{equation:evidence})
and (\ref{equation:likelihood}).
The primary computational task is
evaluating the multidimensional
integral to evaluate the model-averaged likelihoods,
although as both the quasar and star models only have two parameters,
a simple quadrature was sufficient
(and ensured the complete repeatability that would not
be provided by Monte Carlo techniques).

%%%%%%%%%%%%%%%%%%%%%%%%%%%%%%%%%%%%%%%%%%%%%%%%%%%%%%%%%%%%%%%%%%%%%%%%%%%%%%%%

%s4 #&#
\section{Results}
\label{section:results}

The above Bayesian selection method was applied
to the sample of $\sim\!10^4$ DQ candidates described in Section~\ref{section:selection},
giving a value of $P_{\mathrm{q}}$ for each source.
Most of the apparently reasonable candidates
(based on having colours in the
$i$, $Y$and $J$ bands consistent with those expected for
DQs)
had $P_{\mathrm{q}}\ll1$ and could immediately be rejected
from further consideration.
In general, the reason for this was that the measured fluxes,
while being consistent with being a DQ,
were also reasonably consistent with being an ordinary star
(e.g., if the noise level was high),
in which case the high prior probability of being a star
is the dominant factor.
A small fraction of
the candidates have $P_{\mathrm{q}}\simeq1$
and remain viable;
a requirement that
$P_{\mathrm{q}}\geq0.1$
was made to define a concrete
selection criterion.
This cut left 1138 candidate DQs with $P_{\mathrm{q}}\geq0.1$.

The next stage of the process was to examine the actual images
of the candidates with $P_{\mathrm{q}}\geq0.1$ to search for any
anomalies\footnote{Anomalies have included the following: diffraction
spikes from nearby
bright stars; bad pixels in the detector; positional offests;
unresolved binary sources; the presence of a nearby galaxy indicating
a probable supernova; etc.}
that might make the measured photometry---and hence
the likelihood---unreliable.
(The approach employed can, again, be understood in terms
of the information content of the data:
the Bayesian ranking utilised all the information in
the catalogued fluxes; the visual inspection made use of information
in the images that was not captured in the catalogues.)
Over $90\%$ of the $P_{\mathrm{q}}\geq0.1$ candidates were hence rejected,
leaving just 107 reliably measured sources with more than a
$10\%$ chance of being a DQ.

This sample of $\sim\!100$ sources was, finally, sufficiently small
that they could all be reobserved, and follow-up measurements were
made of all 107 candidates in at least one of the
$i$, $z$, $Y$ and $J$ bands.
After each new measurement $P_{\mathrm{q}}$ was recalculated and the
candidate discarded if the probability ever fell below 0.1.
In most cases even a single measurement was sufficient
to reject a candidate; in the remainder
$P_{\mathrm{q}}$ tended to increase towards unity as more measurements
were made.
If $P_{\mathrm{q}}\simeq1$
after follow-up observations in (at least) the
$i$, $Y$ and $J$ bands, then a spectrum was obtained
to confirm the interpretation as a DQ (and to obtain its basic properties).
This process is essentially $100\%$ efficient, in the sense
that all sources for which spectra have been obtained
since the Bayesian selection method was adopted have been confirmed
as quasars.
This contrasts quite strongly with more conventional approaches
to this problem in which the number of spectroscopic observations
per discovery was higher by a factor of between
$\sim\!10$ (e.g., \citealt{Fanetal:2001})
and $\ga50$ \citep{glikmanetal:2008}.

In total, six new DQs have been identified from the UKIDSS data
in this way,
and the seven previously known quasars in the UKIDSS
area were successfully recovered
\citep{Mortlocketal:2012a}.
The new discoveries included
ULAS~J1120+0641 \citep{Mortlocketal:2011d},
which is currently the most distant quasar known.
It is seen as it was just $0.75 \unit{billion years}$ after the Big Bang.
As the only bright source known that early in the history of the
Universe, ULAS~J1120+0641 is   the
target of a large number of follow-up science observations,
including by the {\textit{Hubble Space Telescope}}.
ULAS~J1120+0641
is currently
one of the most important astronomical sources known,
and Bayesian methods were critical to its discovery.

%%%%%%%%%%%%%%%%%%%%%%%%%%%%%%%%%%%%%%%%%%%%%%%%%%%%%%%%%%%%%%%%%%%%%%%%%%%%%%%%

\section*{Acknowledgements}
This research would not have been possible without the efforts
of the SDSS and UKIDSS teams.

%%%%%%%%%%%%%%%%%%%%%%%%%%%%%%%%%%%%%%%%%%%%%%%%%%%%%%%%%%%%%%%%%%%%%%%%%%%%%%%%

% \bibliography{references}

% imsref loaded by akundreckaite, 2013-07-22 13:30:34
% imsref loaded by akundreckaite, 2013-07-24 10:06:20

%

% zodis "Acknowledgments" paliekamas pagal autoriu

%suskaldyti doi

\end{document}